\documentstyle[preprint,aps]{revtex}
\begin{document}
\title
{Dyson's Brownian Motion and Universal Dynamics of Quantum Systems.}
\author{Onuttom Narayan}
\address
{Department of Physics, Harvard University, Cambridge, MA 02138\\{\rm and}\\
AT\&T Bell Laboratories, 600 Mountain Avenue, Murray Hill, NJ 07974}
\author{B. Sriram Shastry}
\address
{AT\&T Bell Laboratories, 600 Mountain Avenue, Murray Hill, NJ 07974}
\date{May 24nd, 1993}
\maketitle

\begin{abstract}
We establish a correspondence between the evolution of the distribution of
eigenvalues of a $N\times N$ matrix subject to a random Gaussian perturbing
matrix, and a Fokker-Planck equation postulated by Dyson. Within this
model, we prove the equivalence conjectured by Altshuler et al between
the space-time correlations of the Sutherland-Calogero-Moser system in
the thermodynamic limit and a set of two-variable correlations for
disordered quantum systems calculated by them. Multiple variable correlation
functions are, however, shown to be inequivalent for the two cases.
\end{abstract}
\pacs{}
\draft

In a series of recent papers, an interesting generalization of the problem
of eigenvalue statistics for complex quantum systems has been introduced by
Altshuler, Simons and coworkers\cite{simon}. They consider the change in the
positions of the eigenvalues in response to an external perturbation; after
an appropriate normalization of the perturbing potential, they show that the
evolution of the eigenvalues of the system as a function of the strength of
the perturbation is universal. By treating this extra parameter (the strength
of the perturbation) as a time-like coordinate, they conjecture that
this problem is identical to the ground state {\it dynamics\/} of an integrable
one-dimensional interacting many body quantum model, the Sutherland
model\cite{sutherland}. The ground state {\it equal time\/} correlations of
this latter model are known to correspond to the eigenvalue statistics at
(any) fixed strength of the perturbation for an appropriate choice of
the coupling constant, depending
on the universality class of the perturbation (real, complex or symplectic).
If the equivalence between the two systems is valid, the full time dependent
correlation functions of this model are then known from Ref.\cite{simon}, and
constitute a dramatic progress in our understanding of this many body problem.
Although the complete spectrum of an integrable many-body problem can often be
found, the time-dependent correlations involve matrix elements, and the
above calculation is one of the very few for a non-trivial many body system.

In this paper, by considering the
Hamiltonian of the perturbing potential to be a random matrix in an appropriate
ensemble\cite{ensemble}, we show the equivalence of
the  evolution of the eigenvalues with the dynamics of the Sutherland model.
To be more precise, by integrating over all perturbations from
a given ensemble, we show that the eigenvalues evolve according to a
Fokker-Planck equation
that is equivalent (under a Wick rotation) to the Calogero
model\cite{calogero}. In the
thermodynamic limit, the Calogero model has identical bulk properties as the
Sutherland model, provided the constants scale properly with N, the size of
the matrix\cite{sutherland2}; this scaling is shown to be fulfilled, whereby we
establish the
conjecture of Ref.\cite{simon}.

For the evolution of eigenvalues, Simons et al consider a system with a
Hamiltonian $H=H_0+xV$, where $V$ is the perturbation, and $x$ is the
strength of the perturbation. Units are normalized so that the mean level
spacing of the eigenvalues $\epsilon_i$ of $H_0$ is unity, as is the
rms ``velocity'' of the eigenvalues, defined as
$\big\langle(\partial\epsilon_i/
\partial x)^2\big\rangle$. The autocorrelation function of the
energy eigenvalues is defined as
\begin{equation}
k(x;\omega)=\bigg\langle \sum_{ij}
\delta\big(\epsilon-\epsilon_i(\overline x)\big)
\delta\big(\epsilon-\omega-\epsilon_j(\overline x +x)\big)\bigg\rangle-1.
\label{corrln}
\end{equation}
By explicit calculation for
disordered systems, Simons et al provide strong evidence
that $k$ is universal, with a form that depends only on whether $H$ is in
the orthogonal, unitary or symplectic ensemble. Numerical simulations are used
to argue that this is also true for quantum chaotic systems.

For $x=0$, $k$ is known to be the same as the ground state
equal time correlations in the positions of a collection of  $N$ particles
confined to a circle, interacting with the Sutherland\cite{sutherland}
Hamiltonian. Here
we consider the alternative Calogero model\cite{calogero}, in which the
particle positions
are allowed to range over $(-\infty,\infty)$, with a confining parabolic
potential:
\begin{equation}
H_C=-\sum_i{{\partial^2}\over{\partial\lambda_i^2}}+\beta(\beta/2-1)\sum_{i>j}
{1\over{(\lambda_i-\lambda_j)^2}}+{1\over{4a^4}}\sum_i
\lambda_i^2.\label{calogero}
\end{equation}
The width of the confining potential is chosen to give a mean ground state
inter-particle separation in the center of the distribution to be unity.
This requirement yields
\begin{equation}
a^2=2N/\pi^2\beta.\label{wellwidth}
\end{equation}
The coupling constant $\beta$ is equal to 1,2 and 4 when $H$ is in the
orthogonal, unitary and symplectic ensembles respectively.
Under the mapping\cite{foot1}
\begin{equation}
x^2=2it,\qquad \epsilon=\lambda\label{oldnorm}
\end{equation}
Simons et al argue that $k(x;\omega)$ is equal to the corresponding {\it time
dependent\/}
correlation function for the Sutherland model. This correlation function can
be calculated explicitly for $\beta=2$, where it agrees with the expression
obtained by Simons et al. For $\beta=1$ and 4, while a complete calculation
is not possible for the Sutherland model, the asymptotic forms of $k$ for
large and small $t$ can be calculated; these agree with the asymptotics of
the expressions derived for $k(x;q)$\cite{simon} for the orthogonal and
symplectic ensembles respectively.

In this paper we demonstrate that, when $V$ is a Gaussian random matrix from
the appropriate ensemble, the evolution of {\it all\/}  quantities  such as
$k$,
involving only the eigenvalues $\epsilon_i(x)$ at two different values
of $x$, are equal to the corresponding time dependent correlation function
for the Calogero model. However, correlation functions involving quantities
at more than two values of $x$ are in general different, so that the full
 dynamics for the two systems are not identical.

In order to do this, it is more convenient to work with
a Fokker-Planck equation that is equivalent to the
Calogero Hamiltonian. With $P\big(\{\lambda_i\};t\big)=\psi_0
\big(\{\lambda_i\}\big)\psi\big(\{\lambda_i\};t\big)$, where $\psi_0$ is the
ground state wave function for the Calogero model, we obtain
\begin{equation}
{{\partial P\big(\{\lambda_i\};t\big)}\over{\partial \tau}}=\sum_i{{\partial^2
P}
\over{\partial\lambda_i^2}}-\sum_{i,j\neq i}
{{\partial}\over{\partial\lambda_i}}\bigg[{\beta\over{\lambda_i-\lambda_j}}
P\bigg]+\sum_i
{{\partial}\over{\partial\lambda_i}}\bigg[{{\lambda_i}\over{a^2}}
P\bigg] .\label{FP}
\end{equation}
Eq.(\ref{oldnorm}) is then changed to
\begin{equation}
x^2=2\tau,\qquad\epsilon=\lambda.\label{norm}
\end{equation}
Eq.(\ref{FP}) is the Fokker-Planck equation corresponding to the Langevin
dynamics of a
collection of $N$ classical particles at finite temperature, with logarithmic
repulsive pairwise interactions, the Wigner-Dyson Coulomb gas\cite{dyson}.
Due to the repulsion between the particles, no steady state distribution
is achieved without the parabolic confining potential. Correlation functions
that involve the particle positions at two different
times $\tau_1$ and $\tau_2>\tau_1$ can be found by considering the evolution
from a general initial state for a time $\tau_2-\tau_1$, and then
averaging over initial states.

For the quantum system, we modify the parametrization of the disorder strength
from $H(x)=H_0+xV$ to
\begin{equation}
H(x)=H_0\cos(\Omega x)+V\sin(\Omega x)/\Omega.\label{reparametr}
\end{equation}
Here $V$ is taken to be of the form $V=\sum_{r=0}^{\beta-1}V_r e_r$, \`a
l\`a Dyson, where $e_r$ are units of the appropriate algebra, so that
$\langle V_{ij}^2\rangle=\sum_{r=0}^{\beta-1}\langle (V_r)_{ij}^2\rangle
=\beta/2\langle V_{ii}^2\rangle$.
This parametrization has the advantage that when $V/\Omega$ and $H_0$ are
considered to be Gaussian random matrices from the same distribution, the
distribution for $H(x)$
is stationary as a function of $x$. For unit mean eigenvalue spacing, we obtain
\begin{equation}
\langle V_{ii}^2\rangle=2\Omega^2 N/\pi^2\beta,\qquad\langle V_{ij}^2\rangle=
\Omega^2 N/\pi^2.
\end{equation}
 From the additional normalization condition that the rms velocity of the
eigenvalues must be unity, it is easy to see from first order perturbation
theory that $\langle V_{ii}^2\rangle=1$, so that we must choose
\begin{equation}
\Omega=\sqrt{\pi^2\beta/2N}=1/a.\label{Omega}
\end{equation}
Since $\Omega\rightarrow 0$ for large $N$, the change we have made in
parametrizing the disorder  strength is inconsequential for finite $x$.

We first note that, for any perturbation $V$, the evolution of the
eigenvalues $\epsilon_i(x)$ as a function of $x$ can be expressed in the
form of a set of first order differential equations in the continuously
changing eigenbasis of $H$:
\begin{eqnarray}
{{d\epsilon_i}\over{dx}}&=& H'_{ii}\nonumber\\
{{d H'_{ii}}\over{dx}}&=&-\Omega^2\epsilon_i+\sum_{j\neq
i}{{2|H'_{ij}|^2}\over{\epsilon_i-\epsilon_j}}\nonumber\\
{{d H'_{ij}}\over{dx}}&=&\sum_{k\neq i,j} H'_{ik}H'_{kj}\bigg(
{1\over{\epsilon_i-\epsilon_k}}+{1\over{\epsilon_j-\epsilon_k}}\bigg)
+H'_{ij}(H'_{jj}-H'_{ii}){1\over{\epsilon_i-\epsilon_j}}.\label{dyneq}
\end{eqnarray}
Eq.(\ref{dyneq}) can be viewed as a classical Newtonian system, with $N$
degrees of
freedom corresponding to the eigenvalues $\epsilon_i$, and $N+\beta N(N-1)/2$
degrees corresponding to the diagonal and off-diagonal elements of $H'$,
obeying appropriate Poisson Bracket relations.
This classical system can be shown to be {\it integrable\/}\cite{dynamics};
this is because the matrices $H$
and $V$ have a very simple dependence on $x$. The integrability of this system
implies that it is
{\it essential\/} to make assumptions about the nature of the matrix $V$,
either of the form we have made above or otherwise.

It is possible to integrate these equations of motion formally; irreversible
equations in $x$ arise when  $V_{ij}$ are averaged over. In the present work,
we find it convenient to follow a different strategy to obtain a Fokker-Planck
equation for the eigenvalues. Our method follows closely along the lines of
a beautiful proof by Dyson\cite {Dyson1} for the behaviour of
the eigenvalues of a matrix subject to random thermal noise; although our
problem does {\it not\/} have a source of thermal noise, this will be seen
to be unimportant for two time correlation functions.

Since $V$ is taken to be a Gaussian random matrix, for any given $H_0$ the
matrix $H(x)$ has the distribution
\begin{equation}
{\cal P}(H)[dH]=\Big[d\big(H/\sin(\Omega x)\big)\Big]\exp\bigg(-{1\over 2}
Tr\Big[\big\{H(x)-H_0\cos(\Omega x)\big\}^2\Big]
{{\Omega^2}\over{\sin^2(\Omega x)}}\bigg).\label{oscdiff}
\end{equation}
We recognize this distribution as the solution to the equation\cite{Dyson1}
\begin{equation}
{{\partial{\cal P}}\over{\partial \tau_Q}}=\sum_{ij}\bigg[g_{ij}
{{\partial^2{\cal P}}\over{\partial
H_{ij}^2}}+{1\over{a_Q^2}}{\partial\over{\partial H_{ij}}}(H_{ij}{\cal
P})\bigg]
\end{equation}
with the initial condition $H=H_0$, where
\begin{eqnarray}
a_Q&=&1/\Omega\nonumber\\
\tau_Q&=&-a^2_Q\ln\big[\cos(\Omega x)\big]\nonumber\\
g_{ij}&=&\delta_{ij}+(1-\delta_{ij})\beta/2.\label{Qnorm}
\end{eqnarray}

This is the Fokker-Planck equation for
a system in which all the matrix elements $H_{ij}$ undergo {\it independent\/}
Langevin dynamics (starting from inital values $(H_0)_{ij}$) in a parabolic
confining well, with an appropriate width and time coordinate.
The temperature at which the Langevin motion takes place is unity for the
diagonal elements of the matrix and $\beta/2$ for the offdiagonal elements.
Eq.(\ref{oscdiff}) thus implies that, for any given $H_0$, {\it all\/} moments
of the different elements of $H(x)$ will be exactly the same at any $x$
{\it as if\/} the elements of $H$ were moving independently in parabolic
confining
wells at a finite temperature. But for such a thermal motion, as shown by
Dyson\cite{Dyson1}, it is possible to go to the eigenbasis of $H$ at any time,
and obtain to second order in perturbation theory for an infinitesmal
increase in time $\delta\tau_Q$ a simplified form of
Eqs.(\ref{dyneq}):
\begin{equation}
\delta\epsilon_i=\delta H_{ii}+2\sum_{j\neq i}[\delta H_{ij}]^2/(\epsilon_i-
\epsilon_j).
\end{equation}
The normalization of Eq.(\ref{Qnorm}) then implies that
\begin{eqnarray}
\langle\delta\epsilon_i\rangle&=&-{{\epsilon_i}\over{a_Q^2}}\delta\tau_Q
+\beta\sum_{j\neq i} 1/(\epsilon_i-\epsilon_j)\delta\tau_Q\nonumber\\
\langle\delta\epsilon_i^2\rangle&=&2\delta\tau_Q.\label{langevin}
\end{eqnarray}
The distribution of eigenvalues, $P_Q\big(\{\epsilon_i\};\tau_Q\big)$ at any
time $\tau_Q$ then satisfies a Fokker-Planck equation like Eq.(\ref{FP}), with
$\lambda_i$ replaced by $\epsilon_i$, and $\tau$ and $a$ replaced by
$\tau_Q$ and $a_Q$ respectively. But comparing Eqs.(\ref{norm})
to Eqs.(\ref{Omega}) and (\ref{Qnorm}), we see that, in
the large $N$ limit, $a=a_Q$ and $\tau=\tau_Q$, yielding Eq.(\ref{FP}).
Thus the distribution of eigenvalues of the quantum system evolves in
exactly the same way as the distribution of particle positions in a
Wigner-Dyson gas.

Note that we have found that the two distributions evolve in the same way for
{\it arbitrary\/} initial conditions. Quantities such as $k(x;q)$ can be seen
from Eq.(\ref{corrln}) to involve a sum over various moments of $\epsilon$,
weighted suitably and then averaged over {\it equilibrium\/} initial
conditions,
requiring only a weaker equivalence. Under a self-averaging assumption, such
equilibrium averages  will be the same as for a generic choice of initial
conditions taken from the equilibrium distribution.

Since our results are independent of initial conditions, it is also possible
to dispense with the parabolic confining well: although there is no
longer any steady state, one can follow the transient dynamics. This actually
corresponds to the original parametrization of $H=H_0+xV$; although
conceptually slightly subtle, the algebra is actually simpler. Although time
translational
invariance is now broken, for any fixed finite time the correction terms
vanish at large $N$.

The result obtained above has been for explicit averaging over a Gaussian
random perturbing potential. Apart from the issue of self-averaging, which is
relevant for quantum chaos, it is
necessary to verify that the additional non-Gaussian terms in the distribution
of $V$ (properly scaled with $N$) do not affect the result in the large $N$
limit, in order to claim universality.
We hope to return to this problem in the future.

While the equivalence between the time coordinate defined in Eq.(\ref{Qnorm})
and Eq.(\ref{norm}) is true only in the large $N$ limit, the form given in
Eq.(\ref{Qnorm}) is true for {\it any} $N$. From the form of
Eq.(\ref{reparametr}),
we see that $2\pi/\Omega$ is the Poincare recurrence `time' interval in $x$
for all the eigenvalues to return to their initial values. From
Eq.(\ref{Qnorm}) we see that the interval $(0,\pi/2)$ in $x$ is stretched out
to $(0,\infty)$
 in $\tau_Q$, so that the system continues to lose memory of its
initial configuration for all $\tau_Q$, with complete equilibration
achieved only in infinite time.

The ground state correlation function of Eq.(\ref{corrln}) can be expressed
for Gaussian random matrices $H_0$ and $V$ in the closed form expression
\begin{eqnarray}
k(x;\omega)&=& \frac{1}{|\sin^\nu(\Omega x)|}\int d{H}\;dH_0
{\rm Tr}\Big[\delta\big(\epsilon-\omega-H\big)\Big]
{\rm Tr}\Big[\delta\big(\epsilon-H_0\big)\Big]\nonumber \\
& &\exp\bigg\{-{1\over{2a^2\sin^2(\Omega x)}}
{\rm Tr}\big[H_0^2+H^2-2H_0 H\cos(\Omega x)\big]\bigg\}-1,
\end{eqnarray}
with $\nu= N + \beta N(N-1)/2$.
{\it All\/} the dependence on time is explicitly present in
the $\cos(\Omega x)$ and $\sin(\Omega x)$ factors.
It is sometimes
possible to evaluate generalized two-matrix Gaussian integrals\cite{mehta},
and it would be interesting to apply these techniques to calculating the above.

It is important to realize that, while Eq.(\ref{oscdiff}) implies that
all moments of the eigenvalues calculated at any $x$ will be equal to
the corresponding moments of the particle positions of a Wigner-Dyson gas
undergoing Brownian motion, this does {\it not\/} mean that the
motion of the eigenvalues is indeed Brownian. The randomness in the
dynamics of the eigenvalues comes from the matrix $V$, which acts like
{\it quenched\/} disorder.
As a simple illustration of the result of the disorder being quenched,  we
consider the case of $N=1$. Eq.(\ref{dyneq}) yields
$\epsilon(x)-\epsilon(0)=V\sin(\Omega x)/\Omega+\epsilon(0)[\cos(\Omega x)-1]$.
Averaging over $V$,  for small $x$  we find that
\begin{equation}
\Big\langle\big[\epsilon(x_1)-\epsilon(0)\big]
\big[\epsilon(x_2)-\epsilon(0)\big]\Big\rangle
=x_1 x_2+\epsilon^2(0)[\Omega^4 x_1^2 x_2^2/4].\label{3pQ}
\end{equation}
For the Wigner-Dyson gas, on the other hand, for $N=1$ we have a particle
in a parabolic well with thermal noise, so that
\begin{equation}
\Big\langle\big[\lambda(\tau_1)-\lambda(0)\big]
\big[\lambda(\tau_2)-\lambda(0)\big]\Big\rangle
=2\min[\tau_1,\tau_2]+\lambda^2(0)[\tau_1\tau_2/a^4].\label{3pD}
\end{equation}
When $\tau_1=\tau_2$, with $2\tau=x^2$, Eqs.(\ref{3pQ}) and (\ref{3pD})
are identical (since $a=1/\Omega$ from Eq.(\ref{Omega})).
However, when $\tau_1\neq\tau_2$, the two
equations are different; this difference persists even for large $N$.

It is precisely such multiple time averages that are involved in three
point (and higher order) correlation functions. For instance, the
density correlation function that is an extension of Eq.(\ref{corrln}) is
\begin{equation}
k(x_1,x_2;\omega_1,\omega_2)=\bigg\langle\sum_{ijl}
\delta\big(\epsilon-\epsilon_i(\overline x)\big)
\delta\big(\epsilon-\omega_1-\epsilon_j(\overline x +x_1)\big)
\delta\big(\epsilon-\omega_2-\epsilon_j(\overline x +x_2)\big)
\bigg\rangle.
\label{threepoint}
\end{equation}
This involves  various moments of the eigenvalues at {\it two}
values of $x$, $x_1$ and $x_2$ (averaged over initial conditions).
For the Wigner-Dyson gas, the three point correlation functions can be
expressed in terms of the two point functions, since the first measurement
`rezeroes' time. If the two point functions could be calculated for the
eigenvalues with {\it arbitrary\/} initial conditions $\{\epsilon_i(0)\}$,
it would be possible to obtain all higher order correlation functions for
the Wigner-Dyson gas. Alternatively, even with only a result for a
representative
choice of $\{\epsilon_i(0)\}$, it might be possible to extract equilibrium
multiple time correlation functions of the Wigner-Dyson gas, if self-averaging
is
valid.

In this paper, we have proved the equivalence of the dynamics of the
Wigner-Dyson gas (or alternatively, the Sutherland model) with the evolution of
the eigenvalues of a Hamiltonian under a perturbation drawn from a Gaussian
orthogonal ensemble for two time correlation functions, as argued by
Simons, Altshuler et al\cite{simon}. Multiple time correlation functions
are not the same in general for the two systems, because of the difference
between annealed and quenched randomness. It is also possible to prove the
equivalence of the two time velocity-velocity correlation function,
$c(\omega;x)$, that measures the correlations in
the rate of change of the eigenvalues\cite{simon}, by a perturbative method, to
all orders in perturbation theory. (This method can also be used to obtain the
Fokker-Planck equation derived
above.) Similar considerations should apply to variants of the quantum
systems considered here, leading to Dyson's circular ensemble\cite{dyson}.
The details will be published in a longer paper.

\centerline{\bf Acknowledgements}
We thank Boris Altshuler, Pierre Hohenberg, Patrick Lee, Ben Simons and
Bill Sutherland for useful discussions, and Boris Altshuler for stimulating
our interest in this problem. O.N. is supported by
a Junior Fellowship from the Society of Fellows at Harvard University.


\begin{references}
\bibitem {foot1} The normalization that we have chosen here is equivalent
to that of Ref.\cite{simon}, since $\theta\equiv 2\pi\lambda/N$, and we
have defined the Hamiltonian with an overall scale factor of $4\pi^2$.
\end{references}
\end{document}